\documentclass[lettersize,journal]{IEEEtran}
\usepackage{amsmath,amsfonts}
\usepackage{array}
\usepackage{textcomp}
\usepackage{stfloats}
\usepackage{url}
\usepackage{verbatim}
\usepackage{graphicx}
\usepackage{cite}
\usepackage[numbers]{natbib}

\usepackage{caption,nccmath}
\usepackage{subcaption}
\usepackage[titlenumbered,ruled]{algorithm2e}
\usepackage{multirow}
\usepackage{cleveref}
\usepackage{algpseudocode}
\usepackage{xcolor}

\begin{document}


\title{Plug-and-Play Model-Agnostic Counterfactual Policy Synthesis for Deep Reinforcement Learning based Recommendation}

\author{Siyu Wang, Xiaocong Chen,~\IEEEmembership{Student Member, IEEE}, Julian McAuley, Sally Cripps and Lina Yao,~\IEEEmembership{Senior Member, IEEE}
\thanks{S. Wang and L. Yao are with the School 
of Computer Science and Engineering, University of New South Wales, Sydney,
NSW, 2052, Australia. \protect E-mail: siyu.wang5@unsw.edu.au}
\thanks{X. Chen  and L. Yao are also with Data61, CSIRO, Australia.}
\thanks{S. Cripps is with Human Technology Institute, University of Technology Sydney, Australia}
\thanks{J. McAuley is with Computer Science Department, UCSD, USA.}
}


\markboth{Journal of \LaTeX\ Class Files, October 2022}%
{Shell \MakeLowercase{\textit{et al.}}: A Sample Article Using IEEEtran.cls for IEEE Journals}


\maketitle

\begin{abstract}
Recent advances in recommender systems have proved the potential of Reinforcement Learning (RL) to handle the dynamic evolution processes between users and recommender systems. However, learning to train an optimal RL agent is generally impractical with commonly sparse user feedback data in the context of recommender systems. To circumvent the lack of interaction of current RL-based recommender systems, we propose to learn a general Model-Agnostic Counterfactual Synthesis (MACS) Policy for counterfactual user interaction data augmentation. The counterfactual synthesis policy aims to synthesise counterfactual states while preserving significant information in the original state relevant to the user's interests, building upon two different training approaches we designed: learning with expert demonstrations and joint training. As a result, the synthesis of each counterfactual data is based on the current recommendation agent's interaction with the environment to adapt to users' dynamic interests. We integrate the proposed policy Deep Deterministic Policy Gradient (DDPG), Soft Actor Critic (SAC) and Twin Delayed DDPG in an adaptive pipeline with a recommendation agent that can generate counterfactual data to improve the performance of recommendation. The empirical results on both online simulation and offline datasets demonstrate the effectiveness and generalisation of our counterfactual synthesis policy and verify that it improves the performance of RL recommendation agents.
\end{abstract}


\section{Introduction}
\IEEEPARstart{T}{raditional} recommendation systems typically rely on content-based filtering~\cite{lops2011content, son2017content} or collaborative filtering approaches~\cite{basilico2004unifying, ramlatchan2018survey} to predict users' future interests based on their past preferences. However, due to the dynamic nature of users' preferences, solely modeling previous interests may not yield accurate predictions. To address this challenge and capture shifting user interests, the concept of dynamic recommendation emerged as a practical technique to enhance recommendation systems through interactive processes~\cite{chen2018stabilizing, chen2020knowledge, zou2020pseudo, chen2022locality}.
The dynamic recommendation involves the system taking optimal actions at each step to maximize the user's feedback reward. Reinforcement Learning (RL) has been recognized as a promising approach for modeling dynamic recommendation systems as it can effectively learn from users' interactive feedback, enabling it to address the evolving nature of user preferences~\cite{dulac2015deep, zheng2018drn,chen2021survey}.
In RL-based recommender systems, the agent will take action
based on the current state of the environment, interact with it, and then receive the reward.
However, one of the primary challenges faced by RL-based recommender systems is the difficulty in precisely grasping users' preferences and generating suitable recommendations, particularly when limited interaction data is available. In such cases, the recommender system may mistakenly assume a lack of interest from the user, resulting in zero rewards when there is no record of a particular circumstance. Consequently, the system's ability to accurately reflect users' actual preferences may be hampered, potentially leading to reduced user satisfaction.

To address the challenge of data sparsity, there has been a growing interest in incorporating the concept of counterfactuals into recommender systems~\cite{xiong2021counterfactual, yang2021top}. Building counterfactuals based on causal relationships offers a solution to the counterfactual question: "What would the interaction process be if we intervened on some parts of the observational data?" By generating counterfactuals from interventions on observational data, we can effectively supplement the information missing from the original data.
This approach, known as data augmentation through counterfactuals, enables the modeling of both the observational and counterfactual data distributions. Leveraging such augmented data equips the recommendation engine with a powerful tool to gain deeper insights into users' genuine preferences. As a result, the recommender system becomes better equipped to adapt to evolving user interests and provide more accurate and personalized recommendations.

Zhang et al.~\cite{zhang2021causerec} propose to measure the similarity between the representation of each item and the target item and replace the top half of items with the lowest similarity scores to obtain the positive counterfactual user sequence.
Wang et al.~\cite{wang2021counterfactual} design a sampler model that makes minor changes to the user's historical items on the embedding space to implement a counterfactual data augment framework.
However, these approaches, while effective for embedding-based methods, face challenges when applied to RL-based dynamic recommendations. Unlike embedding, the state representation in RL is dynamic and influenced by the agent's actions, encompassing user features, feedback, demographic information, and other details. This dynamic nature of state representation is conceptually different from the static nature of embeddings that contain only users' recent actions. As a result, existing counterfactual generating frameworks are not readily compatible with RL-based dynamic recommendation systems.


\begin{figure*}[h]
    \centering
    \begin{subfigure}[b]{0.45\linewidth}
         \centering
         \includegraphics[width=\linewidth]{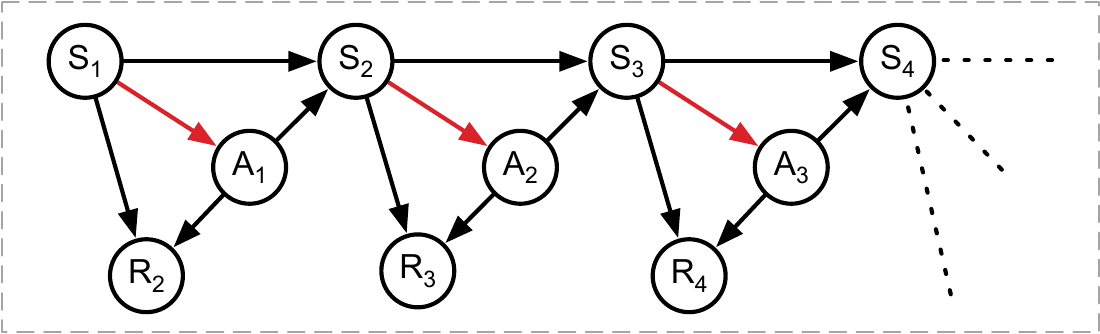}
         \caption{}
         \label{SCM_a}
    \end{subfigure}
    \begin{subfigure}[b]{0.176\linewidth}
         \centering
         \includegraphics[width=\linewidth]{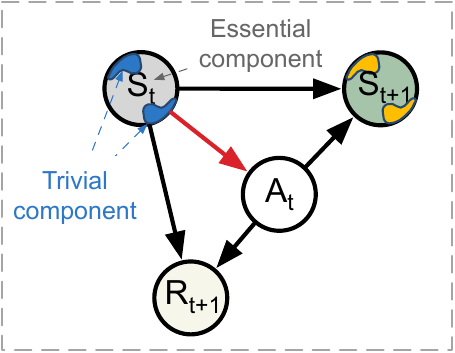}
         \caption{}
         \label{SCM_b}
    \end{subfigure}
    \begin{subfigure}[b]{0.289\linewidth}
         \centering
         \includegraphics[width=\linewidth]{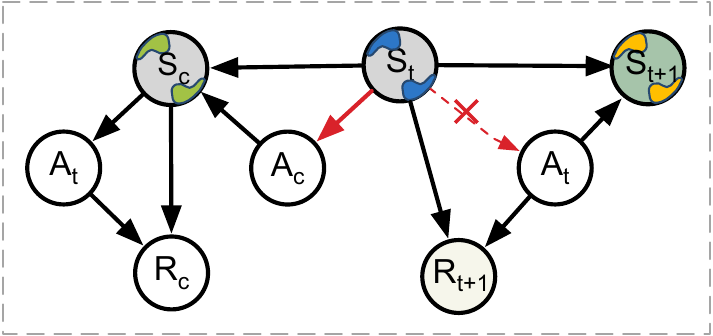}
         \caption{}
         \label{SCM_c}
    \end{subfigure}
  \caption{Causal diagram for MDP. The red edges indicate where we can intervene on. (a) An general MDP instance; (b) one-step MDP to express the possible state changes; (c) Intervening on the action by the $A_t$ to generate counterfactual state $S_c$.}
  \label{SCM}
\end{figure*}

To address these challenges, we propose a novel approach of training a counterfactual synthesis policy to generate counterfactuals tailored to users' dynamic interests in RL-based dynamic recommendations. We identify essential and trivial components in the state representation based on their varying influence on the user's interest. Our objective is to enable the agent to identify optimal actions that selectively change only the trivial components in a state. We treat the modified state as an intervention on the original state to evaluate the causal effect of states on rewards. Since rewards in a recommender system directly reflect users' interests, we can indirectly measure the causal effect of states on users' preferences through their impact on rewards. Consequently, we identify modified components in a state as trivial if their alterations result in a weak causal effect on the reward. This weak causal effect is evident when the intervened reward's distribution closely resembles the initial distribution, indicating a minimal impact on users' interests.
As a result, by minimizing the difference between observational and intervention reward distributions, we can effectively replace a state's trivial components.


Our proposed counterfactual synthesis policy is model-agnostic, making it compatible with any RL algorithm and facilitating its integration with the recommender policy to generate counterfactuals during the interaction process. This approach effectively addresses the data sparsity problem by modeling both counterfactual and observational interaction data.
In summary, our main contributions are as follows:
\begin{itemize}
\item We propose a novel approach to generate counterfactuals based on users' dynamic interests in RL-based dynamic recommendations, effectively modeling both observational and counterfactual data distributions to address data scarcity.
\item We introduce the Model-Agnostic Counterfactual Synthesis (MACS) policy, providing two learning methods. The policy can be seamlessly incorporated into any RL algorithm to collaborate with the recommender policy and generate counterfactuals during the interaction process.
\item We theoretically analyze the identification of causal effects in the recommender system and introduce an effective reward to guide our agents in detecting and replacing the trivial components in a state.
\item We conduct extensive experiments on online simulations and offline datasets, demonstrating the applicability of the counterfactual synthesis policy across various RL algorithms and its significant performance improvement in recommender systems.
\end{itemize}


\section{PRELIMINARIES}

\subsection{RL-based Dynamic Recommendation}
The interactive process between users and a recommendation system  achieved by RL can be formally described as training an agent that interacts with an environment, which follows a Markov Decision Process (MDP)~\cite{sutton2018reinforcement}. 

Specifically, the agent interacts with the environment in each step of a discrete-time sequence $t = 0, 1, 2, ..., n$. In each interaction at time $t_i$, the agent initially receives the state representation $S_t \in \mathcal{S}$ from the environment and chooses an action $A_t \in \mathcal{A}(S_t)$ based on the state $S_t$.
The interaction with the environment then yields a reward $R_{t+1} \in \mathcal{R}$ back to the agent, and the environment will enter into the next state $S_{t+1}$. Formally, the process described above can be formulated by a tuple $(\mathcal{S}, \mathcal{A}(S_t), \mathcal{R}, \mathcal{P}, \gamma)$, where:
\begin{itemize}
    \item $\mathcal{S}$: the set of all states, including the initial and terminal state. State representations contain some information of the environment for the agent to make decisions.
    \item $\mathcal{A}(S_t)$: the set of available actions in the state $S_t$. 
    \item $\mathcal{R}:\mathcal{S} \times \mathcal{A} \to \mathbb{R}$ is the set of possible rewards related to user feedback. 
    \item $\mathcal{P}$: $\mathcal{S} \times \mathcal{A} \times \mathcal{S}\to \mathbb{R}$ are the state-transition probabilities.
    \item $\gamma$: discount factor that satisfies $0 \leq \gamma \leq 1$.
\end{itemize}
In RL, 
a policy ($\pi: \mathcal{S} \to \mathcal{A}$) is a mapping from perceived states to actions that the agent would take when in those environment states. Each interaction between the agent and the environment will yield an immediate reward $R_t$. The learning agent aims to select the action to maximise discounted return, which is the sum of discounted rewards it receives over the episode:
\begin{equation}
\begin{aligned}
G_t := & R_{t+1} + \gamma R_{t+2} + \gamma ^2 R_{t+3} + \gamma ^3 R_{t+4} + \cdots  \\ 
= & \sum_{n=1}^{\infty} \gamma^n R_{t+n}, 
\end{aligned}
\end{equation}
where $T$ is the ﬁnal time step of an episode. 

\begin{figure*}[h]
  \centering
  \includegraphics[width=\linewidth]{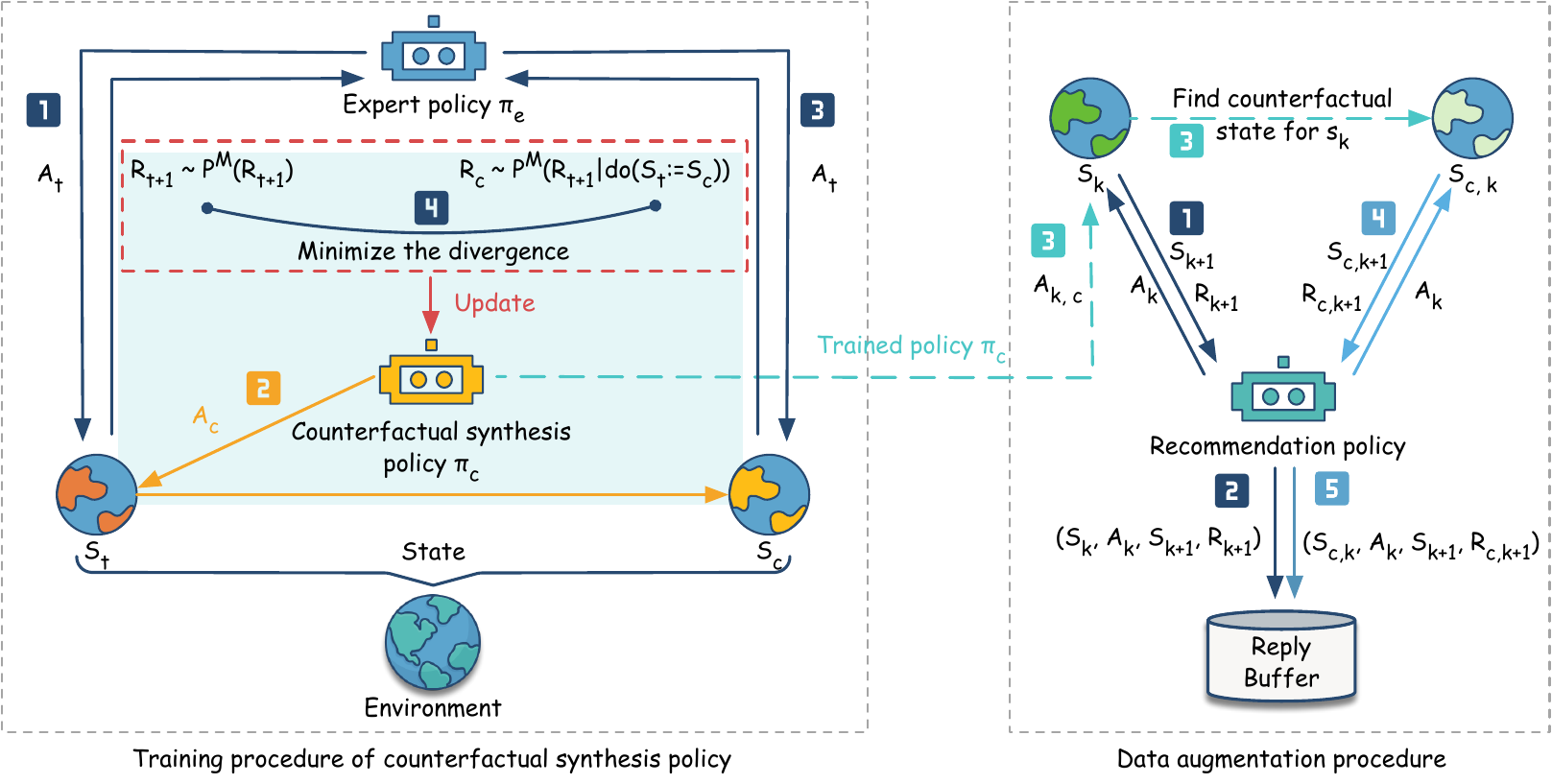}
  \caption{An illustration of the proposed counterfactual synthesis policy. The left part is the schematic of how to train the counterfactual synthesis policy (MACS). The right part is how to apply the counterfactual synthesis policy to other RL algorithms for data augmentation.}
  \label{train}
\end{figure*}
\subsection{Structural Causal Models}
The causal model describes a system by random variables that can be divided into two separate sets: endogenous variables and exogenous variables.
The value of an endogenous variable is determined by the state of the other variables in the system. And in contrast, the value of exogenous variables is determined by factors outside the causal system, meaning that it is independent of other variables in the system~\cite{halpern2020causes}. 

A standard approach to modelling the causal relationships of interacting variables is using structural equations.
For a complex system, Structural Causal Modeling (SCM) is a practical way to describe the causal relationships between variables with a set of structural equations. 
In an SCM, causal relationships are generated by functions that compute variables from other variables. Formally, we refer to these functions as assignments and define SCM as follows~\cite{peters2017elements}:

\vspace{1mm}\noindent\textbf{Definition 1. (Structural Causal Models)}.
\textit{A Structural Causal Model (SCM) $\mathcal{M} := (S, P_U)$ is associated with a directed acyclic graph (DAG) \( \mathcal{G} \), which consists of a collection $S$ of $n$ structural assignments} :

\begin{equation}
\begin{aligned}
X_i := f_i(\boldsymbol{PA}_i, U_i), \:\: i = 1,...,n, 
\end{aligned}
\end{equation}
where $V = \{X_1,...,X_n\}$ is a set of endogenous variables, and $\boldsymbol{PA}_i \subseteq \{X_1,...,X_n\} \setminus \{X_i\}$ represent parents of $X_i$, which are also called direct causes of $X_i$. And $U = \{U_1,...,U_n\}$ are noise variables, determined by unobserved factors. We assume that  noise variables are jointly independent. Correspondingly, $P_U$ is the joint distribution over the noise variables. Each structural equation $f_i$ is a causal mechanism that determines the value of $X_i$ based on the values of $\boldsymbol{PA}_i$ and the noise term $U_i$.

Since the definition of SCM requires its underlying graph be acyclic, each SCM $\mathcal{M}$ models a unique distribution over the endogenous variables $V = \{X_1,...,X_n\}$. Intuitively, SCMs with different causal structures have different distributions. Through intervention on the causal structures, we can obtain intervention distributions that are usually different from the observational one.
We refer the definition of intervention as follows~\cite{peters2017elements}:

\vspace{1mm}\noindent\textbf{Definition 2. (Intervention)}.
\textit{Given an SCM $\mathcal{M} := (S, P_N)$ over $X$, an intervention $\mathcal{I}$ is defined as replacing one or several structural assignments of the SCM $\mathcal{M}$. 
Assume we replace the assignment for $X_k$
by the following expression:}
\begin{equation}
\begin{aligned}
X_k := \widetilde f_k(\widetilde {\boldsymbol {PA}_k}, \widetilde N_k).
\end{aligned}
\end{equation}
Then we say that variable $X_k$ have been intervened on, resulting in a new SCM $\widetilde {\mathcal{M}}$. The corresponding  distribution is changed from the observational distribution $P_X^{\mathcal{M}}$ to the intervention distribution $P_X^{\widetilde {\mathcal{M}}}$, expressed as:
\begin{equation}
\begin{aligned}
P_X^{\widetilde {\mathcal{M}}} := P_X^{\mathcal{M}; do(X_k := \widetilde f_k(\widetilde {\boldsymbol {PA}_k}, \widetilde N_k))}, 
\end{aligned}
\end{equation}
where the operator $do(X_k := \widetilde f_k(\widetilde {\boldsymbol {PA}_k}, \widetilde N_k))$ denotes the intervention that we use to replace the assignment $X_k$. Note that the new noise variables $\widetilde N_k$ are also required to be jointly independent with the original noise variables $N_k$ in the SCM $\widetilde {\mathcal{M}}$.

Another approach to modifying the SCM is to maintain the causal relationship between the variables, but change the distribution of the noisy variables in the causal structure.
In a recommendation system, for example, the counterfactual would deal with the question: "What does the user want to buy if he/she has previously clicked on a different item?". This counterfactual query is based on observed data (the same list of items) but considers a different interventions distribution (clicked a different item). Answering the counterfactual question can be regarded as the process of simulating the causal effect of modifying the actual course of history with minimal change of the causal mechanisms~\cite{pearl_2009}.

\vspace{1mm}\noindent\textbf{Definition 3. (Counterfactual)}.
\textit{Given a SCM $\mathcal{M} := (S, P_N)$ over $X$, a counterfactual is defined as replacing the distribution of noise variables by the noise distribution based on the condition $X=x$, where $x$ are some observations. This can be expressed as:}
\begin{equation}
\begin{aligned}
\mathcal{M}_{X=x} := (S, P_N^{\mathcal{M}|X=x}), 
\end{aligned}
\end{equation}
where $ P_N^{\mathcal{M}|X=x}$ denotes the distribution of the noise variable $N$ when we observe the variable $X=x$ in the SCM $\mathcal{M}$.

\section{Methodology}
\subsection{Problem Formulation}\label{Formulation}
To learn a counterfactual generation policy, we start by formalizing MDP in the semantics of SCMs. The endogenous variables $V$ in SCMs contain the states $\mathcal{S}$, available actions $\mathcal{A}(S_t)$
and the rewards $\mathcal{R}$. The exogenous variables are the noise variables. The causal mechanisms in SCMs are made up of three components: policy $\pi$, state-transition probabilities $P\in\mathcal{P}$ and reward function $f_R$. 
The causal relationship corresponding to the above variables can be illustrated in~\Cref{SCM_a}. 
We assume no confounders in the environment.
The SCMs for MDP can be formulated by the following functions:
\begin{equation}
\begin{aligned}
S_{t+1} := P(S_t, A_t),\:
A_t := \pi_t(S_t),\:
R_{t+1} := f_R(S_t, A_t),
\end{aligned}
\end{equation}

Our work decomposes the state $S_t \in \mathcal{S}$ into two disjoint components, \textit{essential components} $S_t^{ess}$ and \textit{trivial components} $S_t^{tri}$, to identify their different levels of influence on learning users interest representations. We denote the decomposition of state as $S_t = S_t^{ess} \oplus S_t^{tri}$.
As shown in~\Cref{SCM_b},
both the essential and trivial components may be changed when transitioning from state $S_t$ to $S_{t+1}$.
Considering users' dynamic preferences, we assume that not all historical interaction data has a substantial causal effect on inferring users' preferences. That is, only the most recent $m$ interactions are more relevant to users' interests.
Since the rewards in a recommender system reflect users' interests, we measure the causal effect on users' interests through rewards. Formally, we identify the essential components of the state $S_t$ containing essential information of the user's interest. 
Changes to the essential components will result in significant changes to the rewards as well as to the items recommended.
In contrast, the trivial components are the part of the state that are less important for representing the user's interest. Thus 
changing the trivial components have a weak causal effect on the reward in the SCMs. Given the identification of two components, the essence of our work is to find and replace the trivial components in a state.
\Cref{SCM_c} shows that we only change the trivial components in the state $S_t$ to generate counterfactual state $S_c$ by intervening on the action by $A_t$, donated as $do(A(S_t):=A_c)$.
\subsection{Counterfactual Synthesis Policy}
The reinforcement learning method handles problems by considering the interaction of a goal-directed agent with an uncertain environment. It specifies that the agent maximises the total reward it receives over time and adjusts its policy based on its experience. The policy also provides guidance for the agent to choose the optimal action in a given state to maximise overall reward. 
Our goal is to allow the agent to change its policy such that the policy can guide the agent to discover the action to only replace the trivial components in a state. 

Formally, the agent chooses an action $A_c$ as the intervention on the action $A_t$, which can be formulated with $do$ calculus as $do(A_t:=A_c)$. According to the Peal's rules of $do$ calculus~\cite{pearl_2009}, the probability distribution for state $S_{t+1}$ induced after intervention can be calculated by:
\begin{equation}
\label{backdoor}
\begin{aligned}
P^{\mathcal{M}; do(A_t := A_c)}(S_{t+1}) 
&= \sum_{S_t}P(S_{t+1}|do(A_t), S_t) P(S_t|do(A_t)) \\
&= \sum_{S_t}P(S_{t+1}|A_c, S_t) P(S_t) \\
&= E_{S_t}P(S_{t+1}|A_c, S_t), 
\end{aligned}
\end{equation}
which is a special case of the back-door adjustment formula. Hence, we assume that the state variable $S_t$ in the SCM defined in~\Cref{Formulation} satisfies the back-door criterion such that it is sufficient for identifying $P^{\mathcal{M}; do(A_t := A_c)}(S_{t+1})$.

\vspace{1mm}\noindent\textbf{Assumption 1. (Back-Door)}.
\textit{For the SCM $\mathcal{M}$ defined in~\Cref{Formulation} with the corresponding DAG $\mathcal{G}$ shown in~\Cref{SCM_a}, 
the variable $S_t$ satisfies the back-door criterion with regard to the pair of variables $(A_t, S_{t+1})$ because it meets the following criteria:
\begin{itemize}
    \item There is no descendant of $A_t$ in $S_t$.
    \item All paths containing an arrow into $A_t$ between $A_t$ and $S_{t+1}$ are blocked by $S_t$.
\end{itemize}
}

With  Assumption 1, the causal effect of $A_t$ on $S_{t+1}$ is identifiable and is given by the~\Cref{backdoor}. We consider the state $S_{t+1}$ after intervention as the counterfactual state $S_c$ of the state $S_t$, where only the trivial components have been affected, that is:
\begin{equation}
\begin{aligned}
S_c = S_t^{ess} \oplus S_c^{tri} \: \text{and} \: S_c \sim E_{S_t}P(S_{t+1}|A_c, S_t)
\end{aligned}
\end{equation}

Since the agent always learns to maximize its reward, we propose to set up the rewards to evaluate whether only the trivial components have been replaced after intervention on the action. 
Given the SCMs defined in~\Cref{Formulation}, we regard the counterfactual state $S_c$ as an intervention on the state $S_t$ and evaluate the counterfactual state by calculating the causal effect on the reward. Formally, we perform intervention on the state $S_t$ via $do(S_t:=S_c)$, and the $do$ calculus provides us:
\begin{equation}
\label{eq10}
\begin{aligned}
P^{\mathcal{M}; do(S_t:=S_c)}(R_{t+1})
&= \sum_{A_t}P(R_{t+1}|do(S_t), A_t) P(A_t|do(S_t)) \\
&= \sum_{A_t}P(R_{t+1}|S_c, A_t) P(A_t|S_c),
\end{aligned}
\end{equation}
from which we can identify the causal effect of $do(S_t:=S_c)$ on the reward $R_{t+1}$.
If the intervened probability distribution of reward is similar to the original distribution, substituting $S_t$ with $S_c$ has a minor causal effect on reward, indicating that it also has a minor influence on learning the user’s interest. 

To this end,  the reward will take the measurement of the distance between two probability distributions into account. We adopt the KL-Divergence to measure how the intervened reward probability distribution $P^{\mathcal{M}; do(S_t:=S_c)}(R_{t+1})$ is different from the original distribution $P(R_{t+1})$:

\begin{equation}
\begin{aligned}
&D_{KL}(P^{\mathcal{M}; do(S_t:=S_c)}(R_{t+1}) \Vert P^{\mathcal{M}}(R_{t+1}))\\
&= \sum_{r \in R} P^{\mathcal{M}; do(S_t:=S_c)}(R_{t+1}) \log \Big( \frac{P^{\mathcal{M}; do(S_t:=S_c)}(R_{t+1})}{P^{\mathcal{M}}(R_{t+1})} \Big)\\
&= \sum_{r \in R} \Big(\sum_{A_t}P(R_{t+1}|S_c, A_t) P(A_t|S_c)\Big) \cdot \\
& \qquad\quad\qquad\qquad \log \Big( \frac{\sum_{A_t}P(R_{t+1}|S_c, A_t) P(A_t|S_c)}{P(R_{t+1})} \Big)
. \label{eq:kl}
\end{aligned}
\end{equation}


\subsection{Model-agnostic Learning and Integration}

\begin{algorithm}[!ht]
    \SetKwInOut{Input}{input}
    \Input{Initial 
    recommendation critic and actor parameters: $\theta_{\mu^{re}}$ and $\theta_{\phi^{re}}$, replay buffer 
    $D_{re}$\;
    }
    \caption{Training and Application Algorithm for MACS Policy with DDPG under joint method}
    \For{episode = 1, G}{
        Receive initial observation state $s_1$\;
        \For{t = 1, T} {
            \textbf{Stage 1: Train Policy $\pi_{re}$}\;
            Agent following $\pi_{re}$ interacts with the environment to generate transition $(s_t, a_t, s_{t+1}, r_{t+1})$\;
            Store transition $(s_t, a_t, s_{t+1}, r_{t+1})$ in $D_{re}$\;
            \textbf{Stage 3: Counterfactual Generation}\;
            \If{counterfactual synthesis policy $\pi_c$}{
                Agent following $\pi_c$ observe state $s_t$ and select action $a_c$\;
                Execute action $a_c$ and observe counterfactual state $s_c$\;
                Agent following $\pi_{re}$ observe counterfactual state $s_c$ and execute action $a_t$\;
                Observe counterfactual reward $r_{c, t+1}$\;
                Store transition $(s_c, a_t, s_{t+1}, r_{c,t+1})$ in $D_{re}$\;
            }
             Sample minibatch of $\mathcal{N}$ transitions from $D_{re}$ and update $\theta_{\mu^{re}}$ and $\theta_{\phi^{re}}$ by using DDPG algorithm\;
        }
        \textbf{Stage 2: Train Policy $\pi_c$}\Comment{Algorithm 2}   
    }
\end{algorithm}

\begin{algorithm}[!ht]
    \SetKwInOut{Input}{input}
    \Input{Initial counterfactual critic and actor parameters:  $\theta_{\mu^c}$ and $\theta_{\phi^c}$, replay buffer $D_c$, $R_n$, $R'_n$, threshold $\epsilon_1$, temperature $\Delta_1$\; 
    }
    \caption{Training Algorithm for MACS Policy with DDPG under joint method}
        \If{average episode reward $\geq$ $\epsilon_1$}{
            Store the current recommendation policy\;
            \For{n = 1, T}{
                Agent following $\pi_{re}$ observe the current state $s_n$ and select action $a_n$\;
                Execute action $a_n$, observe and store reward $r_n$ in $R_n$\;
                Agent following $\pi_{c}$ observe the current state $s_n$ and select action $a'_n$\;
                Execute action $a'_n$ and observe counterfactual state $s'_n$\;
                Agent following $\pi_{re}$ observe counterfactual state $s'_n$, execute action $a_n$, observe and store intervention reward $r'_n$ in $R'_n$\;
                \If{n\%10}{
                Scale $r_n$ to [0, 1]\;
                $r = r_n + \frac{1}{\text{KL}(R_n || R'_n)}$\;
                Store transition $(s_n, a'_n, s'_n, r)$ in $D_{c}$\;
                Sample minibatch of $\mathcal{N}$ transitions from $D_{c}$ and update $\theta_{\mu^c}$ and $\theta_{\phi^c}$ by using DDPG\;
                }
            }
            Store the current counterfactual policy\;
            $\epsilon_1 = \epsilon_1 + \Delta_1$\; 
        }
\end{algorithm}

Given the above-described reward function, it is possible to develop a policy that meets our goal of identifying a counterfactual state for the current state in which only the trivial components have been replaced.
The key to establishing the above-described reward function is obtaining the reward probability distribution of the observational data and the intervened reward probability distribution.
To this end, we utilize an additional policy to help extract both the observational and intervened reward probability distribution.
Our proposed counterfactual synthesis policy is a model-agnostic approach due to its key point on the reward function.
Thus, it can be easily implemented in current RL-based algorithms to achieve the training process and perform data augmentation. 
We design \textbf{two strategies} for implementing the architecture as mentioned above: the first one is learning the counterfactual synthesis policy assisted by the expert demonstrations. The second one is joint training of the counterfactual synthesis policy and recommendation policy, in which the recommendation policy also serves as the distribution construction policy.
\subsubsection{Learning with Expert Demonstrations}\label{pre}
We introduce a pre-trained policy as the expert policy that uses external knowledge to obtain the observational and intervened reward distribution.
The pre-trained policy can be learned using any RL-based algorithm, and so does our policy.
The idea is motivated by the policy distillation~\cite{rusu2015policy}, in which the student policy is learned by minimizing the divergence between each teacher policy and itself over the dataset of the teacher policy. 
Consider an expert recommendation policy that has learned the knowledge about user interest. It can guide the agent to select an optimal action to get positive feedback from users.
By interacting with the environment, the agent following the expert policy $\pi_e$ will construct the reward distribution over the observational data that reflects user interests. When retaining the same actions, the expert policy $\pi_e$ can also assemble the reward distribution under the intervention of $S_t$ with the~\Cref{eq10}.

To provide specific details, we illustrate the one-time-step learning process with the expert demonstrations architecture in the left part of~\Cref{train}.
Depending on the current state $S_t$ at time $t$, the agent following the expert policy $\pi_e$ takes the action $A_t$ and interacts with the environment to obtain the reward $R_{t+1}$. 
Meanwhile, the counterfactual synthesis agent will apply intervention on the action under the same environment whose current state is $S_t$ to generate a counterfactual state $S_c$.
By replacing the action $A_t$ with the action $A_c$ taken by our training agent, the environment would enter into the counterfactual state $S_c$.
For the expert policy, the generated counterfactual state $S_c$ can be regarded as an intervention on the state $S_t$. Under this intervention, the expert policy can construct the intervened reward distribution and evaluate the causal effect of this intervention on the user interest.
Specifically, we estimate the reward by putting the agent with expert policy $\pi_e$ in the environment, of which the current state is counterfactual $S_c$, to receive the intervened reward $R_c$.

By iterating through the following steps, we can obtain the reward distribution over the observational data $P^{\mathcal{M}}(R)$ and the reward distribution under the intervention $P^{\mathcal{M}}(R')$.
Then the counterfactual synthesis policy is trained to minimize the KL-Divergence between the observational and intervention reward distributions. As a result, we introduce an effective reward formula for the counterfactual synthesis policy. This reward formula serves as a crucial guide for our agents, aiding them in accurately detecting and replacing the trivial components within a state:
\begin{equation}
\begin{aligned}
r 
&= \frac{1}{D_{KL}(P^{\mathcal{M}}(R') \Vert P^{\mathcal{M}}(R)) + \epsilon}
, \label{eq:reward}
\end{aligned}
\end{equation}
where $\epsilon$ is a small constant to prevent the denominator from being zero. In order to maximise the total reward, the agent will select the optimal action to make the intervened probability distribution similar to the original one. Thus, the agent will adjust its policy to achieve only replacing the trivial components in a state.

Once learned, the counterfactual synthesis policy can be used in any RL-based dynamic recommendation to collaborate with the recommendation policy for counterfactual synthesis. 
The procedure follows the right part in~\Cref{train}.
Formally, based on the current state $S_k$, the trained policy $\pi_c$ is used to find and replace the trivial components in  $S_k$ to synthesize counterfactual state $S_{c, k}$. 
The recommendation policy $\pi_{re}$ receives the counterfactual state $S_{c, k}$ but perform the action $A_k$ based on the state $S_k$ to get the reward $R_{c, k+1}$.
Then the counterfactual transition $(S_{c,k}, A_k, S_{k+1}, R_{c, k+1})$ can be put into the replay buffer. 
The generated counterfactual can provide additional information about the user's interests for the recommendation policy, as it is based on changing states. This can also be regarded as exploration for recommendation policy while retain the current level of exploitation.

\subsubsection{Joint Training}

The goal of joint training is to combine the training processes for the recommendation policy $\pi_{re}$ and the counterfactual synthesis policy $\pi_{c}$. To do this, the training procedure is divided into three stages. The conventional recommendation policy training process, which can be based on any RL algorithm, is the first stage. We begin by training the recommendation policy until we store a policy with an average episode reward greater than a certain threshold. 
The stored policy can be used to build the observation and intervention reward distribution. The second stage, training the counterfactual synthesis policy, can then begin. 
This stage follows the approach outlined in~\Cref{pre} with a slight modification to the reward formula as presented in~\Cref{eq:reward}.
\begin{equation}
\begin{aligned}
r &= R'_{t+1} + \frac{1}{D_{KL}(P^{\mathcal{M}}(R') \Vert P^{\mathcal{M}}(R)) + \epsilon}
, \label{eq:reward2}
\end{aligned}
\end{equation}
where $R'_{t+1}$ represents the normalization of $R_{t+1}$. The rationale behind including the normalization of $R_{t+1}$ lies in the joint training process. At the initial stage, the recommendation policy stored for assisting the training of the counterfactual synthesis policy may not be sufficiently optimal, resulting in transitions with lower rewards. By incorporating the normalization of $R_{t+1}$ into the reward function, our counterfactual synthesis policy can learn from a larger set of well-performing recommendation transitions. This approach ensures that the reward takes into account both the transition performance and the similarity of the two reward distributions.
Then the third stage is to apply the learned policy $\pi_c$ to the first stage to construct the counterfactual together with the policy $\pi_{re}$. 
In this approach, the recommendation policy aids policy $\pi_c$ training, while the trained policy $\pi_c$ provides counterfactuals to supplement the transition in the recommendation policy's reply buffer, assisting the recommendation policy in comprehending the users' interests. 
We raise the threshold each time we finish training the policy $\pi_c$. If the recommended policy meets the new threshold, we will use it to restart the second stage to improve the policy $\pi_c$. 
The above-described three stages can use any RL algorithm as an underlying framework. We take the DDPG as an example and present the process in Algorithm 1, in which the training objective can be indicated as minimising the loss function:
\begin{align}
& L(\theta_\mu, \mathcal{D}) = \nonumber\\
& \mathop{E}_{(s, a, s', r) \sim \mathcal{D}} \Big[\Big(\big(r+\gamma(\mu_{\theta^{targ}_\mu}(s', \phi_{\theta^{targ}_\phi}(s'))\big) - \mu_{\theta_\mu}\big(s, a\big)  \Big)^2 \Big],
\end{align}
where $\mathcal{D}$ is a set of mini-batch of transitions $(s, a, s', r)$ for $s \in S$, $a\in A(s)$, $r \in R$, and $s' \in S^+$ ($S^+$ is $S$ plus a terminal state). $\theta_\mu$ and $\theta_\phi$ are parameters for the critic and actor network, respectively. And $\mu_{\theta^{targ}_\mu}$ represent the target critic network.

\section{Experiments}
\subsection{Experimental Setup}
\subsubsection{Data}
We conduct experiments and evaluate our model in both {\it online} and {\it offline} manners. A public simulation
platform, VirtualTaobao~\cite{shi2019virtual}, is used for online evaluation. The benchmark datasets MovieLens-100k
, MovieLens-1M, BookCross and Douban-Book
are used for offline evaluation. 

\vspace{1mm}\noindent\textbf{Online Evaluation}
\begin{itemize}
\item{\textbf{VirtualTaobao}} mimics a real-world online retail environment for recommender systems. It is trained using hundreds of millions of genuine Taobao data points, one of China's largest online retail sites.
The VirtualTaobao simulator provides a "live" environment by generating customers and generating interactions, in which the agent may be tested with virtual customers and the recommendation system.
It uses the GAN-for-Simulating Distribution (GAN-SD) technique with an extra distribution constraint to produce varied clients with static and dynamic features. The dynamic attributes represent changing interests throughout an interactive process. It also employs the Multi-agent Adversarial Imitation Learning (MAIL) technique to concurrently learn the customers' policies and the platform policy to provide the most realistic interactions possible. Each virtual customer has 3-dimensional dynamic attributes and 11 static attributes, such as age and gender, which is encoded into 88 binary dimensions. The attributes of each item are encoded into a 27-dimensional space.
\end{itemize}

\vspace{1mm}\noindent\textbf{Offline Evaluation}
\begin{itemize}
\item{\textbf{MovieLens}~\cite{harper2015movielens}}. MovieLens-100k and MovieLens-1M are stable benchmark datasets based on user ratings on watching movies on the MovieLens website collected during different periods. Ratings are given on a 5-star scale, with each user having at least 20 ratings. Each user is assigned five features, whereas each movie has 23 features with 19 different genres.

\item{\textbf{Douban-Book~\cite{zhu2020graphical} and BookCrossing~\cite{ziegler2005improving}}} are book datasets which contains user's rating and books information. They come from two different book reading websites and both of them are more sparse than Movelens as they contains more user and item information but less interactions.

\end{itemize}


\subsubsection{Baselines and evaluation metrics}
The proposed Counterfactual Synthesis Policy is model-agnostic that can be employed in various popular RL algorithms. 
Although some priors works also approach counterfactual reasoning by generating counterfactual sequences, their works are designed for supervised learning, which is different from RL. 
Existing RL-based recommendation methods do not have unified state representations, which cannot produce a fair comparison. 
Therefore, we mainly focus on the following RL algorithms as baselines:

\begin{itemize}
    \item \textbf{Deep Deterministic Policy Gradient (DDPG)}~\cite{lillicrap2015continuous}. DDPG is an off-policy method for environments with continuous action spaces. DDPG employs a target policy network to compute an action that approximates maximisation to deal with continuous action spaces.
    \item \textbf{Soft Actor Critic (SAC)}~\cite{haarnoja2018soft}. SAC is an off-policy maximum entropy Deep Reinforcement Learning approach that optimises a stochastic policy. It employs the clipped double-Q method and entropy regularisation that trains the policy to maximise a trade-off between expected return and entropy.
    \item \textbf{Twin Delayed DDPG (TD3)}~\cite{fujimoto2018addressing}: TD3 is an algorithm that improves on baseline DDPG performance by incorporating three key tricks: learning two Q-functions instead of one, updating the policy less frequently, and adding noise to the target action.
    \item \textbf{CTRL}~\cite{lu2020sample}: CTRL is an offline RL-based data augmentation method that utilizes the counterfactural method.
    \item \textbf{TPGR}~\cite{chen2019large} is a model designed for large-scale interactive recommendations, which combines the strengths of reinforcement learning with a binary tree structure.
    \item \textbf{PGPR}~\cite{xian2019reinforcement} is an explainable recommendation model that incorporates knowledge awareness and utilizes reinforcement learning techniques.
\end{itemize}

\begin{figure}[h]
  \centering
  \includegraphics[width=0.8\linewidth]{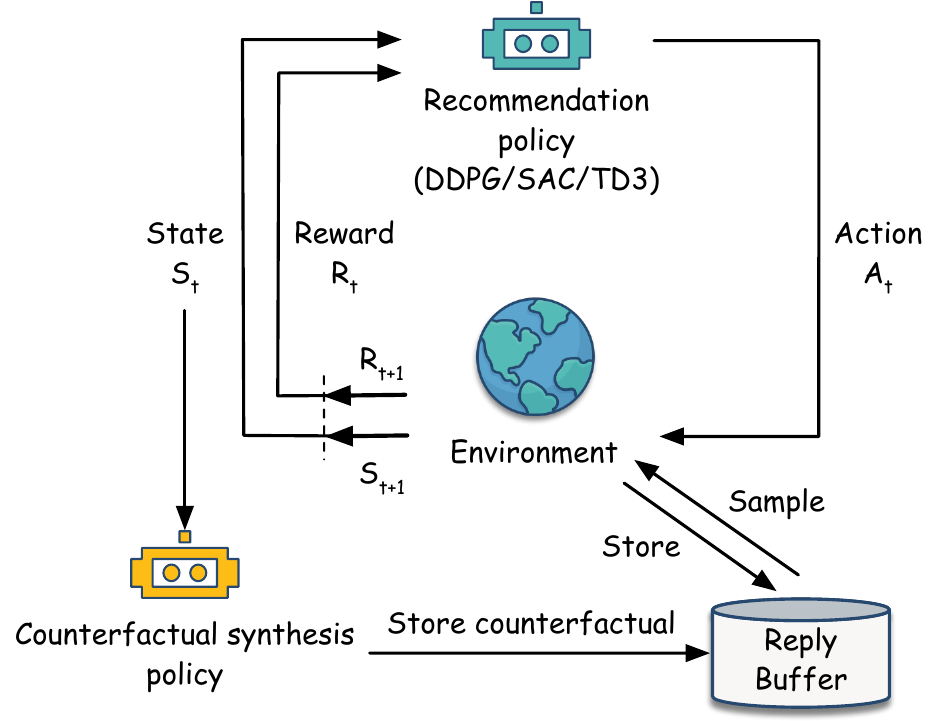}
  \caption{An illustration of how the proposed counterfactual synthesis policy integrate with DDPG/SAC/TD3 in experiment.}
\end{figure}

In terms of evaluation measures, click-through rate is the primary indicator used by VirtualTaobao. For dataset evaluation, three widely used numerical criteria are utilised: Precision, Recall, and Accuracy.

\subsubsection{Implementation Details}
In our experiments, we employ the Deep Deterministic Policy Gradient (DDPG) algorithm with default parameters as specified in \cite{lillicrap2015continuous} to construct the expert policy. The policy is trained for a total of 1,000,000 episodes, and we save the policy with the best performance as our expert policy.
For the training of our proposed method, we set the learning rate to $10^{-4}$ for the actor network and $10^{-3}$ for the critic network. The discount factor $\gamma$ is set to 0.95, and we use a soft target update rate $\tau$ of 0.001. The hidden size of the network is set to 128, and the replay buffer size is set to $10^6$.
Regarding the joint training method, we use a threshold  $\epsilon_1$ of 40\% of the maximum reward (i.e., 4 out of 10) to determine whether the recommendation policy can be stored for training the counterfactual synthesis policy. The temperature parameter used is set to 10\% of the maximum reward to determine whether the stored recommendation policy should be updated.
For all baseline methods, including the joint training method, we adopt the parameter settings as outlined in stable baselines3\footnote{https://stable-baselines3.readthedocs.io/en/master/}.
The experiments are implemented using PyTorch and conducted on a server with two Intel (R) Xeon (R) CPU E5-2697 v2 CPUs, 6 NVIDIA TITAN X Pascal GPUs, and 2 NVIDIA TITAN RTX GPUs.

\subsection{Overall Comparison}
\begin{figure*}[h]
     \centering
     \begin{subfigure}[b]{0.325\linewidth}
         \centering
         \includegraphics[width=\linewidth]{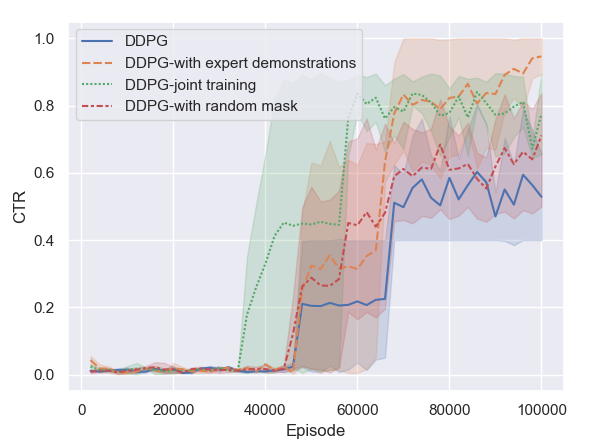}
         \caption{}
         \label{a}
     \end{subfigure}
     \begin{subfigure}[b]{0.325\linewidth}
         \centering
         \includegraphics[width=\linewidth]{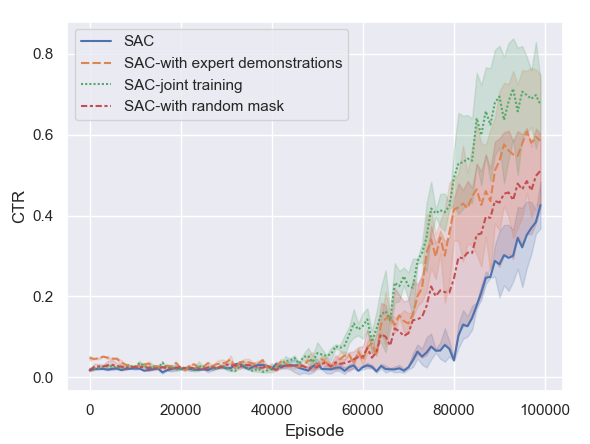}
         \caption{}
         \label{b}
     \end{subfigure}
     \begin{subfigure}[b]{0.325\linewidth}
         \centering
         \includegraphics[width=\linewidth]{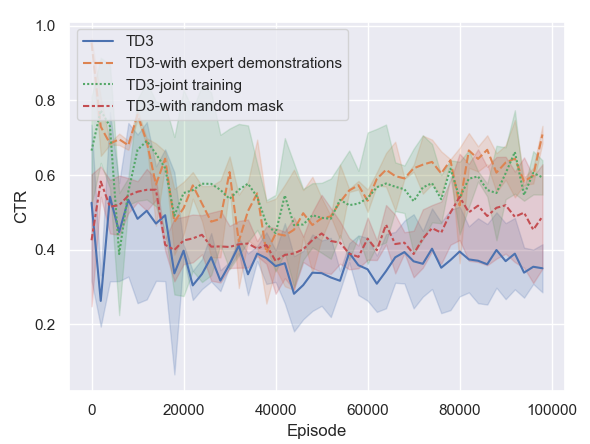}
         \caption{}
         \label{c}
     \end{subfigure}
        \caption{Overall comparison result in online setting between the baselines, baselines with counterfactual synthesis policy learned with expert demonstrations and joint training of baselines and our policy: (a) DDPG as the baseline; (b) SAC as the baseline; (c) TD3 as the baseline. The ablation study results are also included.}
        \label{fig:Comparison}
\end{figure*}

{\bf Online Experiment.} The overall comparison results conducted on the VirtualTaobao platform are depicted in~\Cref{fig:Comparison}. In a nutshell, we find that both our approaches of learning through expert demonstrations and joint training achieve significant improvements over the chosen baselines.
Among the baselines, DDPG achieves the best performance. It obtains a stable policy around the 70,000 episodes but suffers more considerable variance than others. 
The learning pace of SAC is slow. At the end of 100,000 episodes, SAC does not finish learning and does not reach a plateau. One probable explanation is that SAC utilizes a stochastic policy that introduces extra noise to the agent. 
TD3 initially receives a better policy but does not maintain it. The downward trend and the fluctuation may be ascribed to the delayed update parameter mechanism.

Applying a policy learned through expert demonstrations to DDPG shows a similar rising tendency as DDPG, shown in~\Cref{a}. However, with our policy, DDPG rises dramatically with each ascent. This may be due to the fact that the counterfactual transactions generated by our policy assist the recommendation policy in better learning the dynamic interests of users.
The counterfactual synthesis policy trained by expert demonstrations assists DDPG in rising more steadily than others and achieving the highest CTR. 
The joint training method helps DDPG rapidly grow and discovers a good policy around 60,000 episodes. However, we can also observe a fluctuation in the last section. 
One of the possible reasons for the different performances of the two methods is that the counterfactual synthesis policy is frozen under the expert demonstrations setting while continuously optimising under the joint training scenario.

As shown in~\Cref{b}, the performance of SAC has a significant improvement when using our policy learned with either approach. The joint training procedure even assists SAC in reaching a plateau of roughly 90,000 episodes. Although the expert demonstration method drops dramatically near the 80,000 episodes, it rapidly returns and continues to increase.
Because the TD3 employs a special strategy to force agent to conduct exploration at the beginning of training, ~\Cref{c} shows that all three lines start with high CTRs. We can also observe that the expert demonstrations approach initially achieves a very high CTR.
Moreover, TD3 utilising our strategy learned from the two approaches displays a similar tendency. They all descend to a relatively stable level after a defined number of steps at the start, which can be attributed to the fact that TD3 changes the policy less frequently and introduces noise into the target action for target policy smoothing.

\begin{table*}[!h]
\centering
\caption{Performance comparisons of baselines, baselines with counterfactual synthesis policy learned with expert demonstrations and joint training of baselines and our policy on the MovieLens datasets, Douban-Book and BookCrossing datasets. "-E" indicates that the counterfactual synthesis policy is learned with expert demonstration, "-J" indicates a joint training approach, and "-CTRL" indicates the CTRL augmentation method. The best results are highlighted in bold and the second best ones use the symbol *.}
\label{Offline}
\begin{tabular}{c|ccc|ccc}
\hline
\multirow{2}{*}{} & \multicolumn{3}{c|}{MovieLens-100k}                                      & \multicolumn{3}{c}{MovieLens-1M}             \\
                  & Recall                 & Precision              & Accuracy               & Recall                 & Precision              & Accuracy               \\ \hline
DDPG              &  0.4611±0.0091   &   0.4182±0.0053       &   0.4512±0.0312      &  0.7440±0.0045        &  0.4310±0.0023          &  0.5820±0.0028         \\
DDPG-E            & \textbf{0.8361±0.0698} & 0.5105±0.0048*         & 0.7285±0.0462*         & 0.7994±0.0189*         & 0.4324±0.0031*         & \textbf{0.8035±0.0171} \\           
DDPG-J            & 0.7755±0.0020*          & \textbf{0.5111±0.0009} & \textbf{0.7468±0.0178} & \textbf{0.8033±0.0266} & \textbf{0.4329±0.0042} & 0.7846±0.0285*         \\
DDPG-CTRL & 0.7644±0.0055 &  0.5108±0.0010 & 0.7322±0.0202 & 0.7555±0.00301 & 0.4325±0.0028 & 0.6822±0.0111\\ \hline
SAC               & 0.6899±0.15            & 0.499±0.0079           & 0.6266±0.1002          & 0.7149±0.0438          & 0.4103±0.0086          & 0.7016±0.0557          \\
SAC-E             & \textbf{0.7849±0.0861} & 0.5102±0.0025*         & \textbf{0.6951±0.0559} & 0.7974±0.0616*         & \textbf{0.4257±0.0084} & \textbf{0.7751±0.0215} \\
SAC-J             & 0.7795±0.0041*         & \textbf{0.5118±0.0017} & 0.6912±0.0604*         & \textbf{0.8055±0.0493} & 0.4222±0.0088*         & 0.7663±0.0323*         \\ 
SAC-CTRL & 0.7022±0.0422 & 0.5023±0.0044 & 0.6423±0.0068 & 0.7554±0.0502 & 0.4188±0.0045 & 0.7124±0.0058\\ \hline
TD3               & 0.6822±0.1352          & 0.5001±0.0043          & 0.623±0.092            & 0.7034±0.0546          & 0.4201±0.0055          & 0.7121±0.0585          \\
TD3-E             & \textbf{0.7706±0.0785} & 0.5068±0.0045*         & 0.6827±0.0521*         & \textbf{0.7885±0.0047} & 0.4315±0.0029*         & \textbf{0.8077±0.0107} \\
TD3-J             & 0.7541±0.0253*         & \textbf{0.5074±0.0051} & \textbf{0.698±0.0443}  & 0.7711±0.0087*         & \textbf{0.4338±0.0083} & 0.7805±0.0139*         \\
TD3-CTRL & 0.7042±0.0912 & 0.5044±0.0049 & 0.6433±0.088 & 0.7424±0.056 & 0.4298±0.0123 & 0.7566±0.0856\\ \hline
TPGR & 0.3758±0.0026 & 0.3242±0.0077 & 0.3698±0.0026 & 0.6889±0.0088 & 0.3827±0.0108 & 0.5023±0.0067 \\
TPGR-E & \textbf{0.5028±0.0109} & 0.4012±0.0023* & 0.4622±0.0109 & 0.7238±0.0102* & 0.4521±0.0099* & 0.6128±0.0078* \\
TPGR-J & 0.4928±0.0077* & \textbf{0.4023±0.0053} & \textbf{0.4827±0.0098} & \textbf{0.7431±0.0147} & \textbf{0.4721±0.0192} & \textbf{0.6244±0.0099}\\
TPGR-CTRL & 0.4827±0.0088 & 0.3898±0.0048 & 0.4728±0.0100* & 0.7027±0.0098 & 0.4438±0.0102 & 0.6024±0.0076\\\hline
PGPR & 0.4252±0.0047 & 0.3988±0.0043 & 0.4128±0.0088 & 0.6927±0.0091 & 0.4023±0.0044 & 0.5122±0.0088\\
PGPR-E & 0.5201±0.0067* & 0.4728±0.0137* & 0.5077±0.0089* & \textbf{0.7524±0.0028} & 0.5023±0.0042 & 0.6024±0.0077*\\
PGPR-J & \textbf{0.5301±0.0033} & \textbf{0.4777±0.0098} & \textbf{0.5192±0.0099} & 0.7422±0.0047* & \textbf{0.5111±0.0048} & \textbf{0.6088±0.0045}\\
PGPR-CTRL & 0.4927±0.0069 & 0.4533±0.0062 & 0.4928±0.0047 & 0.7322±0.0046 & 0.4765±0.0044 & 0.5837±0.0075\\\hline
\end{tabular}

\begin{tabular}{c|ccc|ccc}
\hline
\multirow{2}{*}{} & \multicolumn{3}{c|}{Douban-Book}                                      & \multicolumn{3}{c}{BookCrossing}             \\
                  & Recall                 & Precision              & Accuracy               & Recall                 & Precision              & Accuracy               \\ \hline
DDPG              &  0.4611±0.0091   &   0.4182±0.0053       &   0.4512±0.0312      &  0.0744±0.0045        &  0.0531±0.0023          &  0.0582±0.0028         \\
DDPG-E            & 0.5322±0.0556* & 0.4922±0.0042*         &    0.5124±0.0120*      &  0.0872±0.0042*       &  0.0582±0.0010      & 0.0690±0.0013* \\
DDPG-J            &  \textbf{0.5612±0.0023}         & \textbf{0.5001±0.0044} & \textbf{0.5201±0.0077} & \textbf{0.0902±0.0012} & \textbf{0.0601±0.0012}  & \textbf{0.0702±0.0023}     \\
DDPG-CTRL & 0.5021±0.0229 & 0.4533±0.0238 & 0.4872±0.0420 & 0.0865±0.0037 & 0.0589±0.0029* & 0.0633±0.0035\\ \hline
SAC               &   0.4700±0.0023     &  0.4203±0.0039         &  0.4400±0.0029         &  0.0721±0.0029        & 0.0511±0.0039        &  0.0572±0.0022       \\
SAC-E             &  0.5120±0.0044* &    0.4912±0.0022*     &  0.5011±0.0045* & 0.0801±0.0011*  & 0.0561±0.0023* & 0.0630±0.0019* \\
SAC-J             &  \textbf{0.5312±0.0011}        & \textbf{0.5010±0.0040} &  \textbf{0.5244±0.0038}        &\textbf{0.0840±0.0020}  &  \textbf{0.0588±0.0019}     &  \textbf{0.0682±0.0018}      \\
SAC-CTRL & 0.5021±0.0033 & 0.4622±0.0029 & 0.4722±0.0077 & 0.0755±0.0045 & ±0.0522±0.0055 & 0.05922±0.0044\\ \hline
TD3               &  0.4711±0.0051        &  0.4199±0.0019        &   0.4488±0.0022         &   0.0702±0.0019      &  0.0510±0.0019        &   0.0566±0.0033        \\
TD3-E             &0.5188±0.0044*  & 0.4899±0.0051*        & 0.4999±0.0046*        & 0.0811±0.0031* & 0.0599±0.0044*         &  0.0641±0.0029*\\
TD3-J             &  \textbf{0.5488±0.0039}       & \textbf{0.4922±0.0032}  & \textbf{0.5099±0.0019}  & \textbf{0.0839±0.0019 }        & \textbf{0.0600±0.0011} &\textbf{0.0671±0.0030}       \\ 
TD3-CTRL & 0.5102±0.0043 & 0.4622±0.0048 & 0.4622±0.0044 & 0.07522±0.0077 & 0.0544±0.0028 & 0.0599±0.0029 \\ \hline
TPGR & 0.4555±0.0192 & 0.4112±0.0048 & 0.4488±0.0027 & 0.0725±0.0032 & 0.0452±0.0044 & 0.0640±0.0041\\
TPGR-E & 0.5299±0.0065* & 0.4822±0.0087* & \textbf{0.5028±0.0077} & 0.0802±0.0055* & 0.0502±0.0056* & \textbf{0.0688±0.0067}\\
TPGR-J & \textbf{0.5308±0.0098} & \textbf{0.4928±0.0076} & 0.5012±0.0059* & \textbf{0.0812±0.0067} & \textbf{0.0523±0.0088} & 0.0659±0.0056*\\
TPGR-CTRL & 0.5201±0.0053 & 0.4722±0.0055 & 0.4823±0.0088 & 0.7877±0.0049 & 0.0487±0.0047 & 0.0660±0.0046\\\hline 
PGPR & 0.4602±0.0099& 0.4182±0.0077 & 0.4460±0.0053 & 0.0699±0.0011 & 0.0393±0.0012 & 0.0733±0.0013\\ 
PGPR-E & 0.5422±0.0120* & 0.4827±0.0088* & \textbf{0.4667±0.0078} & \textbf{0.0765±0.0066} & 0.0455±0.0100* & \textbf{0.0788±0.0044}\\
PGPR-J & \textbf{0.5502±0.0098} & \textbf{0.4876±0.0095} & 0.4657±0.0055* & 0.0755±0.0046 & \textbf{0.0479±0.0078} & 0.0765±0.0032* \\
PGPR-CTRL & 0.5338±0.0076 & 0.4800±0.0056 & 0.4599±0.0088 & 0.0743±0.0067 & 0.0423±0.0067 & 0.0746±0.0066\\\hline
\end{tabular}
\end{table*}

\vspace{0.2cm}
\noindent{\bf Offline Experiment.} The comparison results of employing our policy with baselines are listed in~\Cref{Offline}. On both datasets, the statistical results show that using our policy learned with any approach clearly outperforms the compared underlying models, DDPG, SAC, and TD3.
In a nutshell, The results demonstrate the efficacy of our model-agnostic counterfactual synthesis policy in enhancing any RL-based recommendation model. It is worth mentioning that expert demonstration learning and joint training approaches generate comparable progress across all three baselines.
We attribute the improved prediction result to the fact that our policy assists the recommendation policy in understanding users' interests.
The improved performance of RL-based recommendation systems in both online and offline experiments demonstrates the generality of our policy.

We obverse that, the proposed method is generally outperformed than baselines. On MovieLens datasets, the pre-train method may perform better than the joint-train version. One possible explanation could be that the MoveLens datasets are more dense than the other two datasets. The pre-train method can easily learn a good policy under such a dense situation and guide the agent to achieve better performance. Meanwhile, we find that the improvement achieved on book-crossing and Douban-Book is significantly higher than Movelens. The possible reason could be that, on sparse datasets, out augmentation methods can boost the model performance by generating more data points. Moreover, compare with another RL-based counterfactual data augmentation method -- CTRL, the proposed MACS outperform CTRL from both the expert level and joint train level.
In the later section, we will also investigate how different augmentation approaches affect the final performance and show the superiority of the proposed MACS method.

\subsection{Hyper-parameter Study}
\begin{figure}[h]
    \centering
    \includegraphics[width=0.9\linewidth]{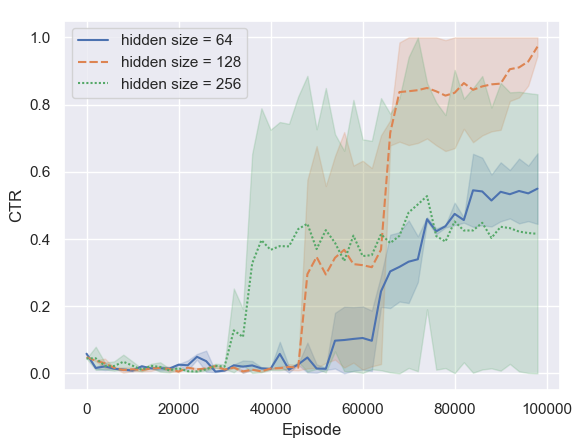}
    \caption{Comparison of performance based on different hidden sizes: 64, 128, and 256.}
    \label{parameter}
\end{figure}

\vspace{1mm}\noindent\textbf{Impact of Hidden Sizes.}
We conduct an ablation study to explore the influence of different hidden sizes on the performance of counterfactual synthesis policies in an online environment. Utilizing the DDPG approach with our policy learned through expert demonstrations as the baseline, we investigate the effect of varying the number of hidden sizes within the range of $[64, 128, 256]$. The comparison results are presented in Figure~\ref{parameter}.
The analysis reveals that the optimal performance is achieved when the hidden size is set to 128. Both excessively small and large hidden sizes negatively impact the performance of our counterfactual synthesis policy. Despite the small standard deviation observed with 64 hidden sizes, it exhibits slower convergence and inferior performance, possibly due to insufficient neurons in the hidden layers to capture essential information. Conversely, an excessive number of hidden sizes results in higher standard deviation, potentially indicating overfitting and hindering convergence.

\begin{figure}[h]
    \centering
    \includegraphics[width=0.9\linewidth]{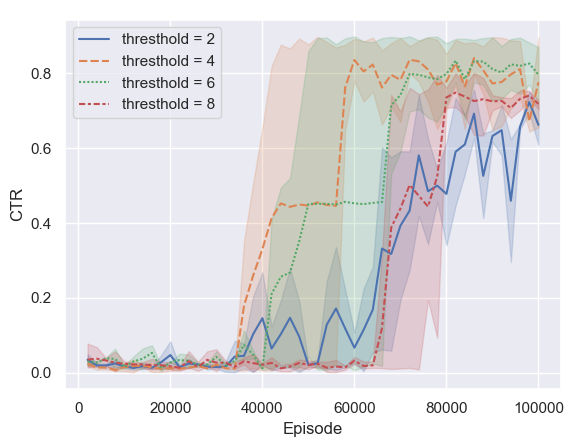}
    \caption{Impact of different threshold values, $\epsilon_1$ ranging from 2 to 8 (with a maximum reward of 10), on the system's performance. The performance metric is evaluated under the joint training method, demonstrating the relationship between threshold choice and the overall system performance.}
    \label{thres}
\end{figure}

\vspace{1mm}\noindent\textbf{Impact of threshold $\epsilon_1$.}
We investigate the performance of the MACS framework under the joint training method across varying threshold values $\epsilon_1$ in the range of 2 to 8, with a maximum reward of 10. The threshold $\epsilon_1$ determines whether the recommendation policy should be stored for training the counterfactual synthesis policy. Results shown in Figure~\ref{thres} indicate that a very small threshold, such as two, leads to unstable performance. This suggests that a suboptimal recommendation policy might negatively affect the learning of the counterfactual synthesis policy. A considerable performance improvement is observed when increasing $\epsilon_1$ to 4. When further increasing $\epsilon_1$ to 6, the performance remains comparable, although the convergence speed is slightly slower, requiring more episodes to achieve the same level of performance as when $\epsilon_1$ equals 4. This is attributed to the counterfactual synthesis policy's delayed training initiation until the average episode reward of the recommendation policy reaches 6. Increasing $\epsilon_1$ to 8 slows down the overall upward trend.
These findings underscore the significance of choosing an appropriate threshold $\epsilon_1$. Setting a threshold that is too small could impede the counterfactual synthesis policy's learning process, while an excessively large threshold might lead to a slower growth trend. Based on our analysis, we recommend training MACS with a threshold of 40\% of the maximum reward.

\subsection{Impact of Counterfactual Augmentation}
In this section, we aim to investigate the effect of the proposed counterfactual data augmentation compared with other data augmentation approaches like random mask. For the random mask, we randomly mask some elements in the state representation that are produced by the environment, then formulate it into a new trajectory and push it into the replay buffer. We are following the same experimental procedure in hyper-parameter study and conduct the experiments on the online environment with those three baselines. The results can be found on~\Cref{fig:Comparison}. We can observe that, the random mask does perform well than the baselines. While it does not achieve a high performance than the proposed methods. One possible explanation is, the random mask can only repeat the existing information to reinforce the learned knowledge without introducing new information. The proposed method utilizes the causal relationship to generate some agent never seen but causally valid trajectory to help learn a better policy. 
\section{Related Work}
\vspace{1mm}\noindent\textbf{DRL-based recommender system.}
Deep reinforcement learning (DRL) has garnered increasing attention in recommender systems research, leveraging the combination of deep learning and reinforcement learning techniques. DRL-based recommender systems model the interaction recommendation process as Markov Decision Processes (MDPs)~\cite{mahmood2007learning}, utilizing deep learning to estimate the value function and tackle high-dimensional MDPs.
Zheng et al.~\cite{zheng2018drn} were among the pioneers to apply Deep Q-Learning in recommender systems, proposing DRN for online personalized news recommendation. To address the dynamic nature of news recommendation, DRN incorporates a DQN to predict potential rewards and considers the frequency of user returns to the app following a recommendation as part of the feedback.
Recognizing the significance of negative feedback in understanding user preferences, Zhao et al.~\cite{zhao2018recommendations} introduced DEERS, which processes positive and negative signals separately at the input layer to avoid negative feedback overwhelming positive signals due to their sparsity.
Chen et al.~\cite{chen2020knowledge} introduced knowledge graphs to DRLs for interactive recommendation, employing a local knowledge network to enhance efficiency.
Hong et al.~\cite{hong2020nonintrusive} proposed NRRS, a model-based approach integrating nonintrusive sensing and reinforcement learning for personalized dynamic music recommendation. NRRS trains a user reward model that derives rewards from three user feedback sources: scores, opinions, and wireless signals.
Chen et al.~\cite{chen2021generative}, instead of predefining a reward function, introduced InvRec, utilizing inverse reinforcement learning to infer a reward function from user behaviors and directly learning the recommendation policy from these behaviors. InvRec employs inverse DRL as a generator to augment state-action pairs, offering a novel approach to the task.

\vspace{1mm}\noindent\textbf{Causal Recommendation.}
Over the past few years, the recommendation domain has seen significant progress in incorporating causal inference techniques. Utilizing causal inference to de-bias training data has been particularly transformative for this field. Schnabel et al.\cite{schnabel2016recommendations} proposed an inverse propensity scores (IPS) estimator to counter selection bias by weighting observations in the recommendation process. Bonner and Vasile\cite{bonner2018causal} introduced a domain adaptation algorithm to leverage biased logged feedback and predict randomised treatment effects using random exposure. Liu et al.\cite{liu2020general} focused on knowledge distillation, presenting KDCRec, a framework to overcome bias problems in recommender systems by extracting information from uniform data. Zhang et al.\cite{zhang2021causal} addressed the confounding influence of popularity bias and devised an inference paradigm to adjust recommendation scores through causal intervention. Additionally, Ding et al.\cite{9737000} proposed CI-LightGCN, a Causal Incremental Graph Convolution method to efficiently retrain graph convolutional networks in recommender systems, employing Incremental Graph Convolution (IGC) and Colliding Effect Distillation (CED) operators to handle model updates with new data while preserving recommendation accuracy. Counterfactual inference has also been applied in the recommender system context for path-specific effects removal\cite{wang2021clicks} and OOD generalisation~\cite{wang2022causal}. Furthermore, an increasing number of works adopt counterfactual reasoning for purposes like providing explanations, enhancing model interpretability, and learning robust representations~\cite{tan2021counterfactual, madumal2020explainable, zhang2021causerec}.

\vspace{1mm}\noindent\textbf{Counterfactual data augmentation in recommendation.}
Counterfactual data augmentation has been recently leveraged as a powerful method in machine learning for alleviating data sparsity, which has seen tremendous success in the fields of neural language processing~\cite{zeng2020counterfactual} and computer vision~\cite{chen2019counterfactual, chen2020counterfactual}.
Lu et al.~\cite{lu2020sample} propose to use counterfactual data augmentation in RL to improve data-efficient. They design a GAN-like adversarial framework to learn a causal mechanism that is used to estimate counterfactual outcomes for alternative actions.
Recent works have introduced this idea in the sequential recommendation domain.
Zhang et al.~\cite{zhang2021causerec} generate counterfactual interaction sequences for sequential recommendation by using similarity function to identify which subsets in the behaviour sequence can be replaced.
Wang et al.~\cite{wang2021counterfactual} propose to generate counterfactual sequences by finding replacement items in the embedding space for sequential recommendation models.

\section{Conclusion}
In this paper, a novel counterfactual synthesis (MACS) policy in plug and play fashion has been proposed based on the casual view of MDP.
The linking reward with the divergence between the observational and intervening reward distributions can guide the agent to find a replacement state with minimal causal effect on users' interest.
The counterfactual synthesis policy is simple to implement in various RL frameworks and works with the recommender agent.
During the interaction between the recommender agent and the environment, our agent provides counterfactual data that takes into account the dynamic users' preferences.
Results show that the counterfactual synthesis policy performs well on different RL frameworks and achieves a considerable improvement for all compared baselines.
\bibliographystyle{IEEEtran}
\bibliography{Full_8}

\end{document}